\documentstyle[twocolumn,aps,epsf,floats]{revtex}
\begin{document}
{\bf Comment on: ``Band Filling and Interband Scattering
Effects in MgB$_2$: Carbon versus Aluminium Doping''}
\vskip 0.2cm

In  a recent Letter  \cite{kortus}  the  authors  claim that the observed decrease of $T_c$ in Al and  C doped
MgB$_2$ samples is mainly due to  the  band  filling  by   the  electron  doping  but  for a fuller understanding
of the different  behavior of the superconducting energy gaps  in Al and  C doped samples  it is necessary  to
also include  an  increased  interband  scattering  for the carbon  doped samples.  We argue  in the following
that this latter statement is misleading  and is based on the assumption that for carbon doping the two
superconducting gaps merge near 10 \% carbon doping: an assumption that contradicts most of the existing
experimental data.
A great deal of experimental effort has been invested into the measurements of the superconducting gaps of 10 \% C
doped MgB2 samples. The thermodynamic measurements of the specific heat in that case  clearly show the large gap
feature and provides    strong    suggestions    of    the   small   gap \cite{ribeiro}. The tunneling
measurements \cite{samuely,schmidt} very nicely do exactly the opposite:  i.e. clearly show the smaller gap and
provide strong indication of the larger gap. Later work, Holanova {\it et al.} has even shown the presence  of
both gaps in a single experiment (see Fig. 2 of Ref. \cite{holanova}).  The theory presented by Kortus et al.
shows the two gaps merging for $T_c$ values very close to that of the 10 \% sample: a prediction that is strongly
inconsistent with these data.
The problem with assuming that the two superconducting gaps merge
near 10 \% carbon doping is even more starkly illustrated by an
examination of the systematic change in gap values with carbon substitution.
Fig. 1 presents the  systematics of the  available data on  the
behavior  of  the  superconducting  energy  gaps upon carbon
doping in  the magnesium diboride.  The solid symbols  are from
the point-contact  tunneling  measurements  by  Holanova {\it et al.}
\cite{holanova,holanova2} and Schmidt et al. \cite{schmidt} and also
from the recent photoemission
spectroscopy by  Tsuda {\it et  al.} \cite{tsuda}. The  open
diamonds
represent  the  gap  values  obtained  by  Gonnelli  et  al.
\cite{gonnelli}
from  the  treatment  of  their  point-contact  spectroscopy
measurements.  All data, with the exception of those from
Gonnelli et  al., are fully  compatible with the  theoretical
model  including only  the band  filling effects.  In this case, the
increasing carbon doping  leads to a decrease of  $T_c$ as well
as to  a proportional decrease  of the both  superconducting
gaps (dashed lines). The dotted curve  is the theoretical prediction of
Kortus  et  al.,  where  beside  the  band  filling an additional
increase in interband scattering  $\gamma _{inter}  =  2000
x$cm$^{-1}$ is
considered. This latter
effect  increases the  value of  the small  gap and together
with the band  filling leads to the indicated merging of
the two gaps.
Without any detailed  analysis of  the original  data of the four groups, the  conclusion can be made that  for
the large gap  there  is  a  perfect  consensus  among  all  presented measurements. As far as the  small gap is
concerned the mean values obtained by Gonnelli {\it et  al.} are positioned at higher energies  than in  another
sets of measurements  but if the error  bars  are  taken  into  account  the  data are not so different, with 
the excpetion of the
point at 18.9  K. The single gap seen by Gonnelli {\it et al.}  on the sample with 13.2  per cent of carbon doped
in MgB$_2$ and $T_c$ of 18.9 K is the central point on which the theoretical  treatment of Kortus  {\it et al.} is
based \cite{gonnelli2},   but  ommiting   all  other  experimental evidence available.
The theoretical calculations of Erwin and Mazin \cite{erwin} have shown that  replacement of boron  by carbon does
not change the  local point  symmetry. In this case,  the interband  scattering should remain  as small, as in the
undoped  MgB$_2$ and  the two superconducting energy  gaps should not be expected to merge, as indeed they do not
(\cite{samuely,schmidt,holanova,holanova2,tsuda}). In  the   measurements  \cite{holanova,holanova2,tsuda} the
carbon doped samples of different forms  (wires, sintered  pellets, polycrystals)  prepared by different methods
were used. Nevertheless, the presence of the two superconducting energy  gaps and their consistent evolution with
T$_c$ without merging has been found directly in the raw data up to the highest carbon concentrations. Ignoring
these experimental data  makes the Letter \cite{kortus} pertaining to a single experimental data point and
contradicting several other established experimental results.
This  work  has  been  supported  by  the Slovak Science and
Technology     Assistance      Agency     under     contract
No. APVT-51-0166.
\begin{figure} [t]
\vspace{-0.5cm}
\begin{center}
\leavevmode
\epsfxsize=7.5cm
\epsffile{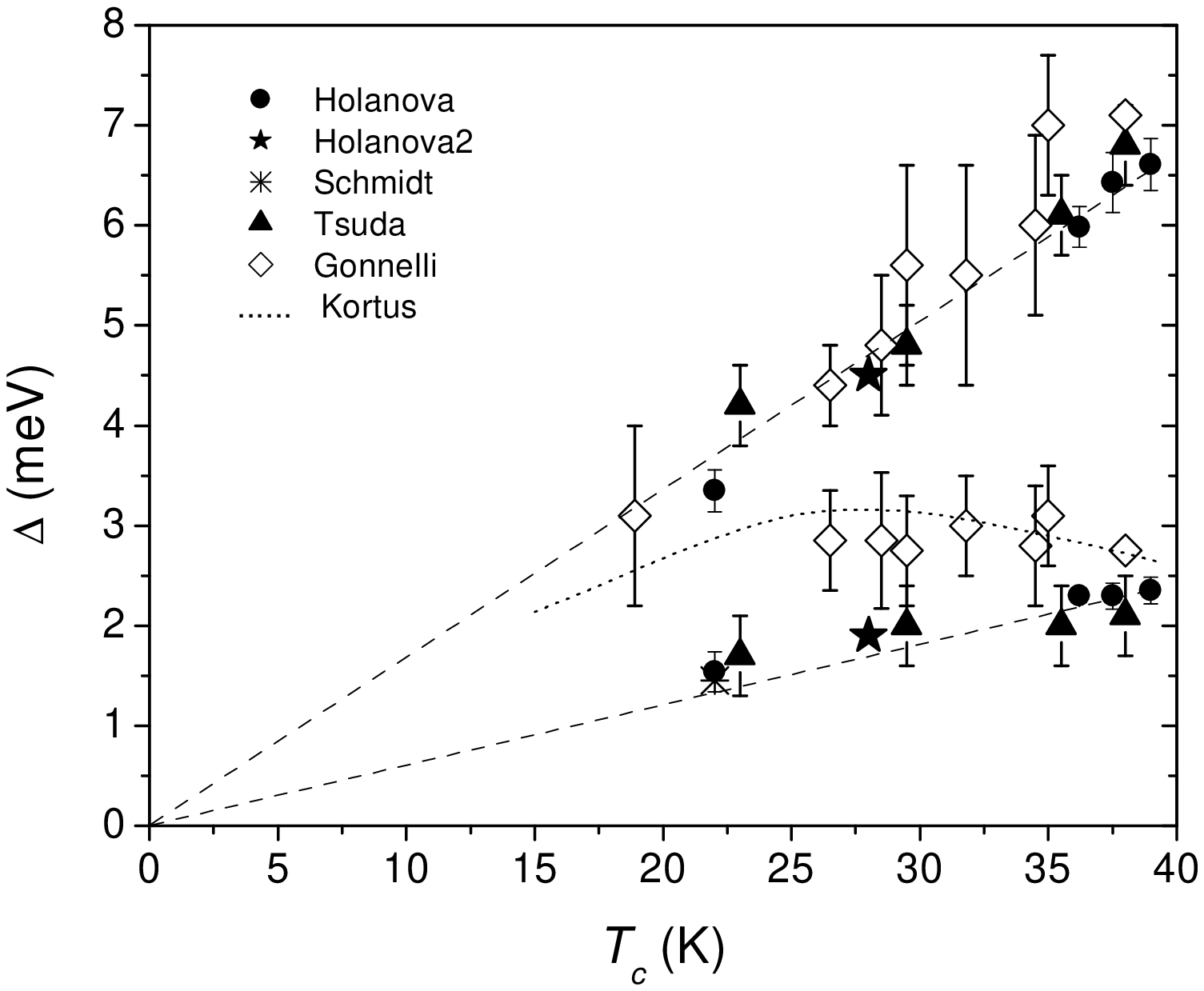}
\end{center}
\caption{Superconducting   energy   gaps  from  different
experiments  (see  text)  on  the  carbon  doped  MgB$_2$ as
a function of $T_c$. Dotted line  - calculation of Kortus et
al. [1]. Dashed lines are guide for the eye.}
\end{figure} 
\vskip 0.3cm
\noindent
P.   Samuely,$^{1}$   P.    Szab\'o,$^{1}$   P.   C.
Canfield,$^{2}$ S. L. Bud'ko$^{2}$
$^1$Centre  of  Low   Temperature  Physics  of  the
Institute of Experimental Physics SAS \& Faculty of Science UPJ\v S,
SK-04353~Ko\v{s}ice,         Slovakia\\
$^2$ Ames Laboratory and Department of Physics and Astronomy,
Iowa State University, Ames, IA 50011 USA

\end{document}